\documentclass[]{emulateapj}
\usepackage{epsfig}
\usepackage{floatflt}

\shorttitle{Tracing Multiple Generations of AGN Feedback in the Core of Abell 262}
\shortauthors{Clarke et al.}

\def\lsim{\mathrel{\rlap{\lower4pt\hbox{\hskip1pt$\sim$}}
    \raise1pt\hbox{$<$}}}                % less than or approx. symbol
\def\gsim{\mathrel{\rlap{\lower4pt\hbox{\hskip1pt$\sim$}}
    \raise1pt\hbox{$>$}}}                % greater than or approx. symbol

\begin{document}

\title{Tracing Multiple Generations of AGN Feedback in the Core of Abell 262}

\author{T.\ E.\ Clarke\altaffilmark{1,2}, E.\ L.\ Blanton\altaffilmark{3}, C.\ L.\ Sarazin\altaffilmark{4}, L.\ D.\ Anderson\altaffilmark{3}, Gopal-Krishna\altaffilmark{5}, E.\ M.\ Douglass\altaffilmark{3}, and Namir E.\ Kassim\altaffilmark{1}}
\altaffiltext{1}{Naval Research Laboratory, 4555 Overlook Ave. SW, Code 7213, Washington, DC 20375, USA}
\altaffiltext{2}{Interferometrics Inc., 13454 Sunrise Valley Drive, Suite 240, Herndon, VA 20171, USA}
\altaffiltext{3}{Institute for Astrophysical Research, Boston University, 725 Commonwealth Ave., Boston, MA 02215, USA}
\altaffiltext{4}{Department of Astronomy, University of Virginia, P.~O.\ Box 400325, Charlottesville, VA 22904-4325, USA}
\altaffiltext{5}{NCRA-TIFR, Pune University Campus, Pune 411 007, India}

\begin{abstract}
We present new radio and X-ray observations of Abell 262. The X-ray
residual image provides the first evidence of an X-ray tunnel in this
system while the radio data reveal that the central radio source is
more than three times larger than previously known. We find that the
well-known cluster-center S-shaped radio source B2 0149+35 is
surrounded by extended emission to the east and south-west. The
south-western extension is co-spatial with the X-ray tunnel seen in
our new $Chandra$ images while the eastern extension shows three
clumps of emission with the innermost coincident with a faint X-ray
cavity. The outer two eastern radio extensions are coincident with a
newly detected X-ray depression.  We use the projected separation of
the emission regions to estimate a lower limit of $\tau_{rep}=28$ Myr
to the outburst repetition timescale of the central AGN. The total
energy input into the cluster over multiple outburst episodes is
estimated to be $2.2\times 10^{58}$ ergs, more than an order of
magnitude larger than previously thought. The total AGN energy output
determined from our new observations shows that the source should be
capable of offsetting radiative cooling over several outburst
episodes.
\end{abstract}


\section{Introduction}

The thermal gas in the cores of dense, relaxed galaxy clusters has a
cooling time that is often shorter than the age of the system
\citep{fn77,cb77}. In the absence of heating, this gas should cool
through X-ray emission and be deposited within the cooling radius in
the form of dense molecular clouds. Over the lifetime of the cluster,
the accumulated mass could reach as much as $10^{11}-10^{12}
M_{\odot}$. Detailed observations over a wide range of the
electromagnetic spectrum fail to detect such large quantities of cool
gas deposited in cluster cores. Optical observations reveal star
formation which reaches only a few percent of the predicted cooling
rates in most systems \citep{mo89,c98,c99}. High resolution X-ray
spectroscopic observations with $XMM-Newton$ show a lack of evidence
for gas cooling below roughly one third of the maximum temperature
\citep{peterson01,tamura01,kaastra01,s02,peterson03}. The lack of
evidence of significant quantities of cool X-ray gas implies that the
mass accretion rates to low temperatures are $\sim$ 10 times lower
than previously estimated \citep{mn07}. One possible origin for this
discrepancy is that there is a heat source in the intracluster medium
(ICM) that stops significant amounts of gas from cooling to low X-ray
temperatures \citep[see review of heating mechanisms
  by][]{conroy08}.

A promising candidate for the source of heat energy in galaxy clusters
is the supermassive black hole residing in the core of the central
dominant galaxy. \citet{burns90} found that 71 percent of cD galaxies
in cooling cores showed evidence of radio outbursts associated with
the central active galactic nucleus (AGN), while only 23 percent of
non-cooling core cD galaxies revealed radio emission. The high spatial
resolution of the Chandra X-ray Observatory has revealed the
ubiquitous presence of cavities associated with radio outbursts in the
cores of dense cooling core clusters \citep[see compilations
  by][]{birzan,df04,r06}. The radio lobes inflated by the central AGN
appear to displace the X-ray gas in the cluster core creating the
depressions in the X-ray emission. These depressions provide a good
means of estimating a lower limit on the mechanical energy input into
the ICM by the AGN outburst. In many cases, the cavities are
associated with radio emission from the current AGN outburst which can
be traced at frequencies of 1400 MHz and higher. In some cases,
however, there are ghost cavities that have been detected in the X-ray
emission but have no radio counterpart at frequencies of 1400 MHz or
higher. These cavities are generally found at projected radii much
larger than the scale of the currently active radio lobes and they are
thought to be the result of past outburst activity. In several
individual sources, such as Perseus \citep{fabian02}, Abell 2597
\citep{clarke05}, and Abell 4059 \citep{clarke07}, radio observations
reveal that the X-ray ghost cavities (or tunnel in the case of Abell
2597) are filled with low frequency radio emission, while in the case
of Hydra A \citep{lane04,nulsen05} an X-ray follow-up of the large low
frequency outer radio lobes revealed the presence of a giant X-ray
cavity.

In several cooling core systems, X-ray observations reveal the
presence of multiple sets of cavities. The most spectacular example of
a multiple cavity system is seen in the Perseus cluster
\citep{hans93,fabian00,fabian06}. Deep $Chandra$ observations of this
system reveal the presence of at least five individual cavities which
are thought to be the remnants of past outburst episodes. Soft X-ray
observations of the Virgo cluster reveal a series of filaments which
trace regions that are thought to be buoyant bubbles from a series of
small outbursts from the central AGN \citep{forman07}. In the case of
the cavity system in Hydra A, \citet{wise07} conclude that the six
observed cavities may be the result of either continuous outflow or a
series of outbursts from the central AGN. In all cases, the presence
of these multiple cavity systems provides new insight into the
lifecycle of the central AGN.

In this paper we present detailed radio observations of the central
radio source in Abell 262. The cD galaxy in Abell 262 (NGC 708) is
host to a double lobed radio source B2 0149+35. Previous radio
observations of this source revealed that it is a fairly weak FR I
radio galaxy \citep{Parma86} with an `S' shaped morphology and a total
linear extent of $\sim$ 19 kpc. $Chandra$ X-ray observations of Abell
262 by \citet{Blanton04} showed a cavity on the eastern side of the
core which was filled by 1400 MHz radio emission. Our analysis of an
elliptical model-subtracted $Chandra$ residual image discovered the
presence of an X-ray tunnel running from the radio core to radii well
beyond the western edge of the \citeauthor{Parma86} source
(\S~\ref{sect:chan_anal}). We present detailed multi-frequency radio
observations of this source which we interpret as evidence of multiple
outburst episodes. Using the results of these new observations, we
discuss the impact of the AGN outbursts on the surrounding ICM.

Throughout the paper we adopt the cosmological parameters $H_o$ = 71
km s$^{-1}$ Mpc$^{-1}$, $\Omega_\Lambda=0.73$, and $\Omega_m=0.27$. At
the redshift of Abell 262 ($z=0.0163$) this cosmology corresponds to a
scale of 0.327 kpc/\arcsec.

\section{Observations}

Low frequency radio observations of Abell 262 were undertaken with the
National Radio Astronomy Observatory's (NRAO) Very Large Array (VLA)
operating at two frequencies near 325 MHz and the Giant Metrewave
Radio Telescope (GMRT) operating at frequencies near 235 and 610
MHz. VLA data obtained at frequencies near 1400 and 4500 MHz are also
included in the analysis presented here. Table~\ref{tbl:radio_obs}
provides the details of the radio observations discussed in this
paper.

\subsection{VLA Data} 

VLA radio observations of B2\ 0149+35 were taken at frequencies around
1400 and 330 MHz with the VLA in A configuration. At 330 MHz these
observations (obtained 2003 July 27) provide a resolution of $\sim$
6\arcsec, while at 1400 MHz (obtained 2000 December 19) the
resolutions is $\sim$ 1.5\arcsec. The source 3C48 was observed as a
calibrator for the 330 MHz observations and was used for bandpass,
flux, and phase calibration of the target. The 330 MHz data were
obtained in spectral line mode to permit excision of radio frequency
interference (RFI) and avoid problems of bandwidth smearing. The 1400
MHz A configuration data were taken in continuum mode and used 3C147
as a flux calibrator and 0201+365 as a phase calibrator. These
observations were supplemented by VLA archival observations at 1400
MHz from B and C configurations (taken in 2005 April 16 and 1997
September 18, respectively) to provide sensitivity to extended low
surface brightness emission. The B configuration 1400 MHz data was
taken in spectral line mode and used 3C48 as bandpass, flux, and phase
calibrator. The C configuration observations were obtained in
continuum mode and used 3C48 as a flux calibrator and 0119+321 as a
phase calibrator. Observations at 4500 MHz in the VLA D configuration
on 2000 October 1 were also reduced in order to obtain total flux
information for the integrated spectrum of the source. These
observations were taken in continuum mode and had 3C286 as a flux
calibrator and used 0145+386 as a phase calibrator.

The data were calibrated and reduced with the NRAO Astronomical Image
Processing System (AIPS). Images were produced through the standard
Fourier transform deconvolution method. All data sets were processed
through several loops of imaging and self-calibration to reduce the
effects of phase and amplitude errors in the data. Frequencies near
1400 MHz and below were reduced using the wide-field imaging
capabilities within the AIPS task IMAGR. The two frequencies near 330
MHz were combined together to make the final image. Similarly all
frequencies and configurations for the 1400 MHz data were combined
into the final image.

\subsection{GMRT Data}

Abell 262 was observed with the GMRT on 2004 July 24. The observations
were taken in dual frequency mode providing two single polarization
(Stokes RR) frequencies near 610 MHz and one single polarization
(Stokes LL) frequency near 235 MHz. The 610 MHz observations had a
bandwidth of 16 MHz for each frequency resulting in a total bandwidth
of 32 MHz and a resolution of $\sim$ 6\arcsec. The total bandwidth of
the 235 MHz data was 8 MHz and the resolution is $\sim$ 13\arcsec. All
data were taken in spectral line mode to allow RFI excision and reduce
problems of bandwidth smearing. At both frequencies we used 3C48 as
bandpass, flux, and phase calibrator.

The GMRT data reductions were carried out with AIPS. The GMRT data
near 610 MHz were fairly clear of interference while the 235 MHz data
required a significant effort to flag the RFI. The calibration
followed standard wide-field imaging techniques and both data sets
were self-calibrated through several loops of phase and amplitude
calibration. The two frequencies near 610 MHz were calibrated and
imaged separately and the final combined image was made using the
image plane deconvolution task APCLN within AIPS.

\subsection{$Chandra$ Data} 
\label{sect:chan}

Abell 262 was observed by $Chandra$ on 2006 November 20 (ObsID 7921)
and 2001 August 3 (ObsID 2215) for 111,934 s and 30,305 s
respectively.  For each observation, the center of the cluster was
positioned on the back-illuminated ACIS-S3 detector with the events
telemetered in very faint (VF) mode and an energy filter of $0.1-13$
keV. The cluster centroid was positioned with a 1\arcmin\ offset from
the nominal pointing to avoid node boundaries. The ``level 1'' events
files were processed with CIAO version 3.4 following the standard
procedure for VF mode. Only events with ASCA grades 0, 2, 3, 4, and 6
were included. By analyzing light curves of the two back-illuminated
detectors (ACIS-S1 and ACIS-S3) throughout the observation period, it
was determined that no background flares occurred to contaminate the
data. The data were corrected for hot pixels and cosmic-ray afterglows
using standard techniques.  The resulting cleaned exposure times were
110,674 s and 28,744 s.

A merged image of both pointings (chips ACIS-S2, ACIS-S3, and ACIS-S4)
was created by reprojecting the event files of the older ObsID 2215 to
the WCS reference frame of our new deeper observations (ObsID 7921).
This resulted in a total combined exposure time of 139,418 s.  Energy
was restricted within the range of 0.3-10.0 keV.  Background files
were taken from the blank sky observations of M. Markevitch included
in the CIAO calibration database (CALDB) and reprojected to match both
observations of A262.  The background files for both pointings were
then merged using ObsID 7921 as a reference.  Matching exposure maps
were created for each observation and merged in a similar manner.

\section{X-Ray Analysis}
\label{sect:chan_anal}

\begin{figure}[tb]
\plotone{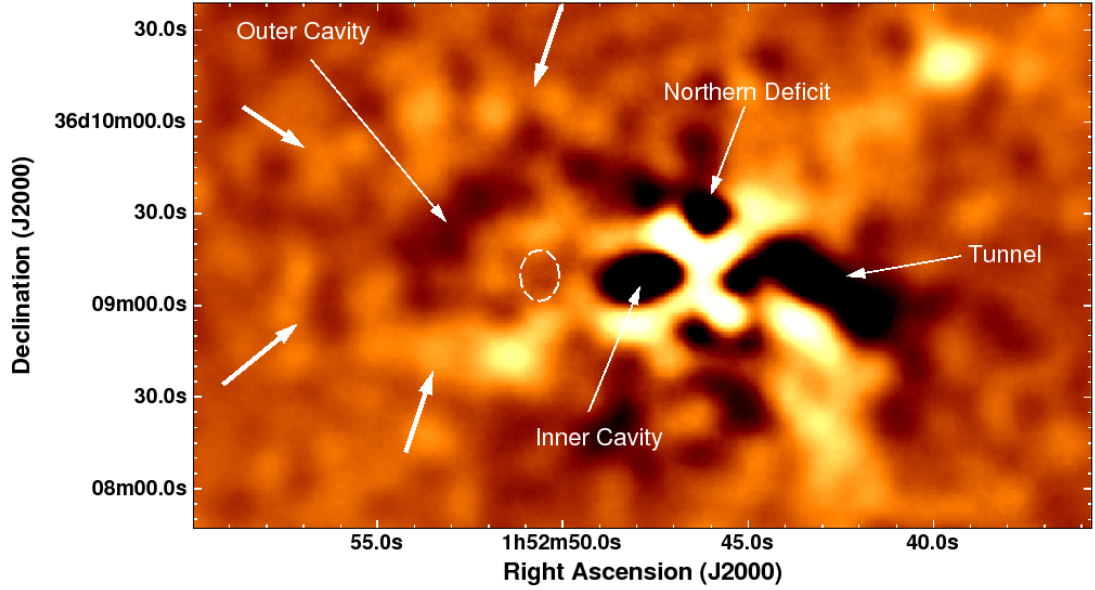}
\caption{Residual $Chandra$ image of the central region (96 $\times$
  56 kpc) of Abell 262, produced by subtracting an elliptical
  isophotal model from the merged $\sigma$=5\arcsec\ Gaussian smoothed
  image in the $0.3-10$ keV band. The eastern X-ray cavity and newly
  discovered western tunnel are the most prominent features of the
  image. To the north of the cluster core there is an X-ray deficit
  located between two bright regions which are coincident with [N
    \textsc{ii}] emission. The dashed ellipse indicates the position
  of the faint X-ray depression discussed in \citet{Blanton04}. The
  new deep residual image also hints to the presence of an extended
  deficit to the east of the dashed ellipse which may be surrounded by
  a faint rim. The labels indicate the deficits while the thick arrows
  mark the rough location of the faint eastern rim.
\label{fig:A262_unsharp}}
\end{figure}

A detailed discussion of the the cluster properties based on the new
merged $Chandra$ data, including spectral fitting, will be presented
in \citet{Blanton09}. For a review of the cluster properties based on
the original short $Chandra$ observations see \citet{Blanton04}. Here,
we concentrate on details of the central region of the cluster. Using
IRAF, we fit a smooth elliptical model to the Gaussian smoothed
($\sigma=5$\arcsec) background and exposure corrected $0.3-10$ keV
merged image of Abell 262. The ellipticity, position angle, and
intensity of the isophotes were allowed to vary about the fixed X-ray
centroid. In Figure~\ref{fig:A262_unsharp} we show the residual image
of the cluster center resulting from subtracting the elliptical model
from the Gaussian smoothed image. In addition to the inner eastern
X-ray hole discussed in \citet{Blanton04}, the residual image also
reveals the first evidence of an X-ray tunnel to the west of the
cluster core.

The inner eastern cavity is significant at the 22 $\sigma$ level and
appears to be nearly completely surrounded by a bright X-ray rim while
the newly discovered western tunnel appears to have only a partial
rim. We have extracted counts in an annulus with inner and outer radii
equal to the distance from the AGN of the inner and outer edge of the
tunnel region, respectively, and excluded the counts from the tunnel
region itself for the annulus counts.  Comparison of the counts in
this region to the counts in the tunnel feature reveals that the
tunnel is significant at the 11 $\sigma$ level. Aside from these two
most prominent features, the residual image also shows evidence of the
faint eastern cavity discussed in \citet{Blanton04} (marked with
dashed circle). Analysis of the new merged $Chandra$ data shows that
this feature is still of low significance compared to X-ray counts at
a similar radius. The residual image also shows a deficit to the north
of the cluster core that is significant at roughly the 8 $\sigma$
level. This deficit is located between two bright X-ray knots that are
associated with [N \textsc{ii}] emission \citep{Plana98} and may
either be a signature of a clumpy ICM or an additional X-ray
cavity. Further to the east of the cluster core there appears to be a
previously unknown extended X-ray deficit surrounded by a faint X-ray
rim. Analysis of this extended complex region shows that it is
significant at only the 2 $\sigma$ level.

\section{Multi-Frequency Radio Analysis}

\subsection{Morphology}

\begin{figure}[tb]
\epsscale{0.7}
\plotone{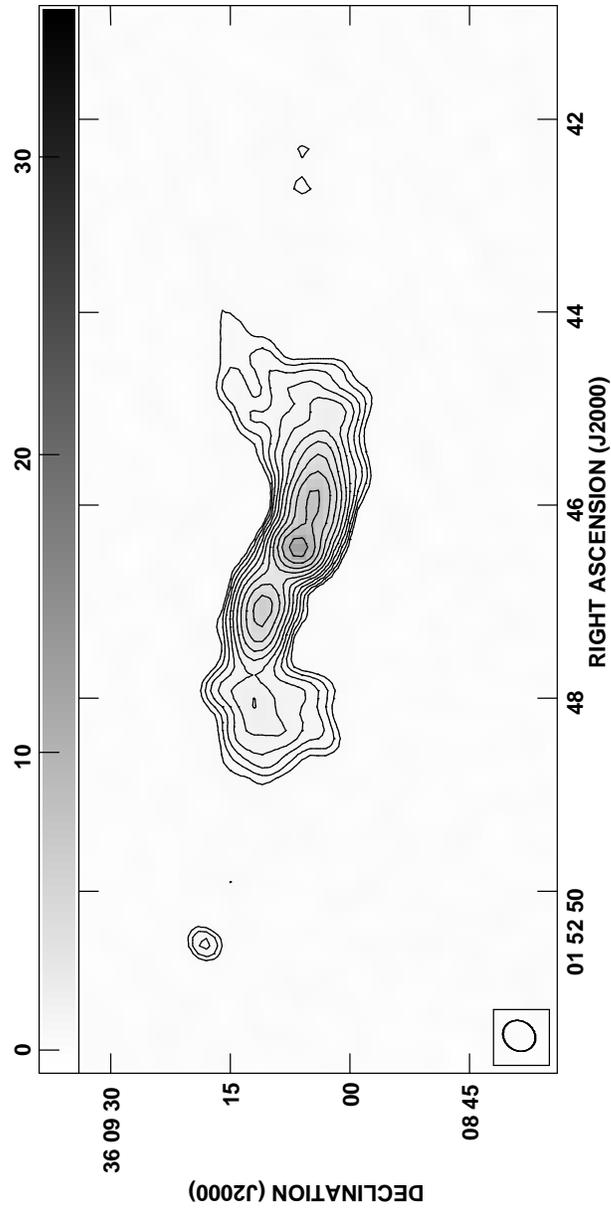}
\caption{NRAO VLA image at 1400 MHz of B2 0149+32. The resolution of
the image is 4\farcs2 $\times$ 3\farcs7 (the beam is shown in the
lower left corner of the image), and the contours increase by
$\sqrt{2}$ from 3$\sigma$ where $\sigma$=0.11 mJy/beam. The bar at the
top of the image shows the total intensity scale in units of
mJy/beam. The compact source at RA(J2000)=01 52 50.5, Dec(J2000)=36 09
18.18 may be an unrelated background source.
\label{fig:VLAB_L}}
\end{figure}

In Figure~\ref{fig:VLAB_L} we show the VLA B configuration 1400 MHz
image of B2 0149+35. The central radio emission contains a compact
core connected to a western extension which may trace the western
radio jet. A bright region to the east of the core may be associated
with the counterjet which would suggest an initial position angle of
$\sim$ 65$^\circ$ (from north to the east) for the jet axis. Beyond
the bright central regions the radio emission extends into the western
and eastern radio lobes. The outer lobes lie at a different position
angle through the core as compared to the bright jet and
counterjet. This variation in the source position angle may indicate
bulk rotation of the surrounding ICM such as suggested by
\citet{gopal} for Centaurus A. Numerical simulations of the impact of
bulk ICM motions on embedded radio galaxies reveal that such motions
can reproduce observed radio galaxy morphologies \citep{heinz}. An
alternative explanation for the structure of B2 0149+35 is that the
position angle of the radio jets has precessed over time as suggested
for clusters such as Perseus \citep{dunn06}, RBS797 \citep{gitti} and
Abell 2626 \citep{wong}.

\citet{Blanton04} showed that the eastern radio lobe is coincident
with the inner eastern cavity seen in the $Chandra$
data. Figure~\ref{fig:VLAB_L} shows that in addition to the S-shaped
radio source, there is a compact radio source located $\sim 17$ kpc
east of the core of B2 0149+35. This source has no X-ray counterpart
and no known optical identification but seems to be located near the
eastern edge of a faint optical source on the Second Generation
Digitized Sky Survey (DSSII) images.  The total radio flux of the
compact source at 1400 MHz is 0.67 $\pm$ 0.17 mJy. Although the low
signal to noise on this source makes the calculation of the spectral
index uncertain, we note that it is consistent with that of typical
extragalactic sources.

\begin{figure}[tb]
\epsscale{1.0}
\plotone{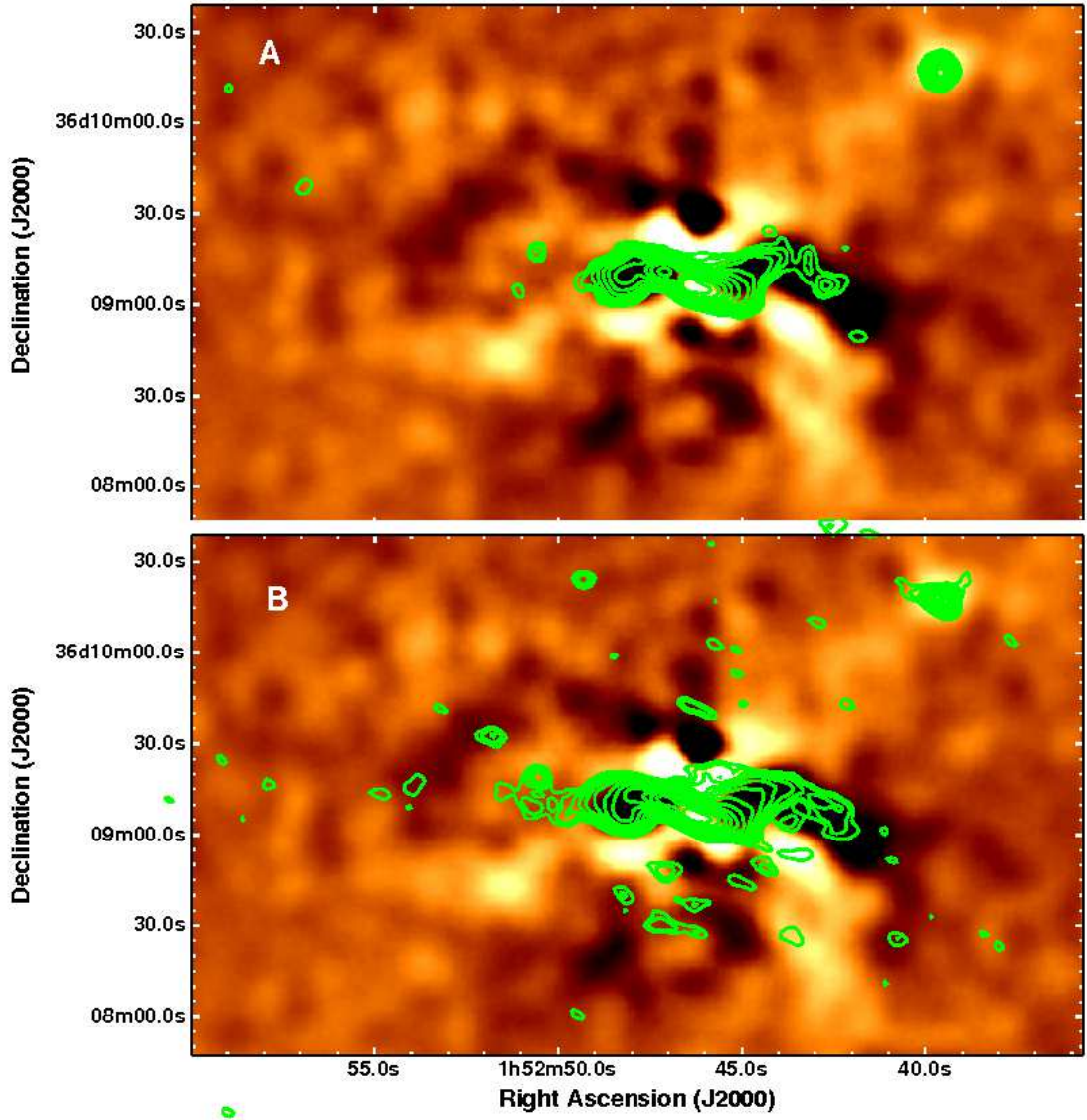}
\caption{{\bf A:} Combined A, B and C configuration VLA image at 1400
  MHz of B2 0149+32 overlaid as contours on the residual $Chandra$
  image (Figure~\ref{fig:A262_unsharp}). The resolution of the radio
  image is 6\farcs0 $\times$ 5\farcs5 and the contours increase by
  $\sqrt{2}$ from 3$\sigma$, where $\sigma$=0.12 mJy/bm. The radio
  emission associated with NGC708 is seen to extend further to the
  west along a tail that runs to the southwest along the X-ray
  tunnel. {\bf B:} Full resolution GMRT 610 MHz radio contours
  overlaid on the residual $Chandra$ image. The radio data have a
  resolution of 5\farcs9 $\times$ 4\farcs6 and the contours increase
  by $\sqrt{2}$ from 3$\sigma$, where $\sigma$=0.11 mJy/bm. The radio
  emission traces further along the X-ray tunnel than seen at 1400 MHz
  and also shows evidence of a radio extension to the east which is
  coincident with the faint X-ray depression discussed in
  \citet{Blanton04}.
\label{fig:mos1}} 
\end{figure}

Combining the A, B, and C configuration 1400 MHz data allows us to
trace more extended radio emission from B2 1049+35 since these data
are sensitive to lower surface brightness
emission. Figure~\ref{fig:mos1}A shows the combined three
configuration data set as contours overlaid on the central region of
the residual $Chandra$ image. We note the presence of bright X-ray
clumps to the north-east and south-west of the radio core at roughly
the projected location where the radio jets change position angle.
Higher resolution radio data of the jets would help determine if there
is a connection between the X-ray clumps and the jet deflection. The
entire radio source appears to be fully confined by the bright X-ray
rims seen in the residual $Chandra$ image.

The radio emission shown in Figure~\ref{fig:mos1}A extends further to
the west and north-west of the radio core than is seen by
\citet{Parma86}. Beyond the north-west emission there is a further
radio extension to the southwest, along the X-ray tunnel. The radio
emission appears to be continuous from the core toward the north-west
to a distance of $\sim$ 34\arcsec\ ($\sim$ 11 kpc) at which point it
appears to break up into two connected clumps to the south-west within
the tunnel. Both clumps reveal extended structure and do not have any
optical identifications; thus they are most likely tracing radio
plasma ejected from the central AGN in NGC 708. At the lowest contour
level (3$\sigma$) the western emission does not reveal an obvious
sharp edge to the source suggesting that we may not be tracing the
full extent of the emission at this frequency. The emission to the
east, beyond the counterjet location, fills the inner eastern X-ray
cavity and appears to have a relatively well defined edge.

Lower frequency full-resolution radio observations at 610 MHz from the
GMRT highlight again the close confinement of the synchrotron emission
by the bright X-ray rims seen in the $Chandra$ image
(Figure~\ref{fig:mos1}B). These new radio data also confirm the
presence of extended radio emission tracing the southwest X-ray tunnel
and show that the emission extends further down the tunnel at this
frequency. They also reveal diffuse emission running from the eastern
radio lobe to the region south of the compact radio source. This
eastern emission is coincident with the location of the faint ghost
cavity mentioned in \citet{Blanton04} and thus supports the
interpretation of this region as an X-ray cavity. The total flux in
the eastern extension (excluding the compact source) is 5.3$\pm$ 0.5
mJy.

\begin{figure}[tb]
\epsscale{0.7}
\plotone{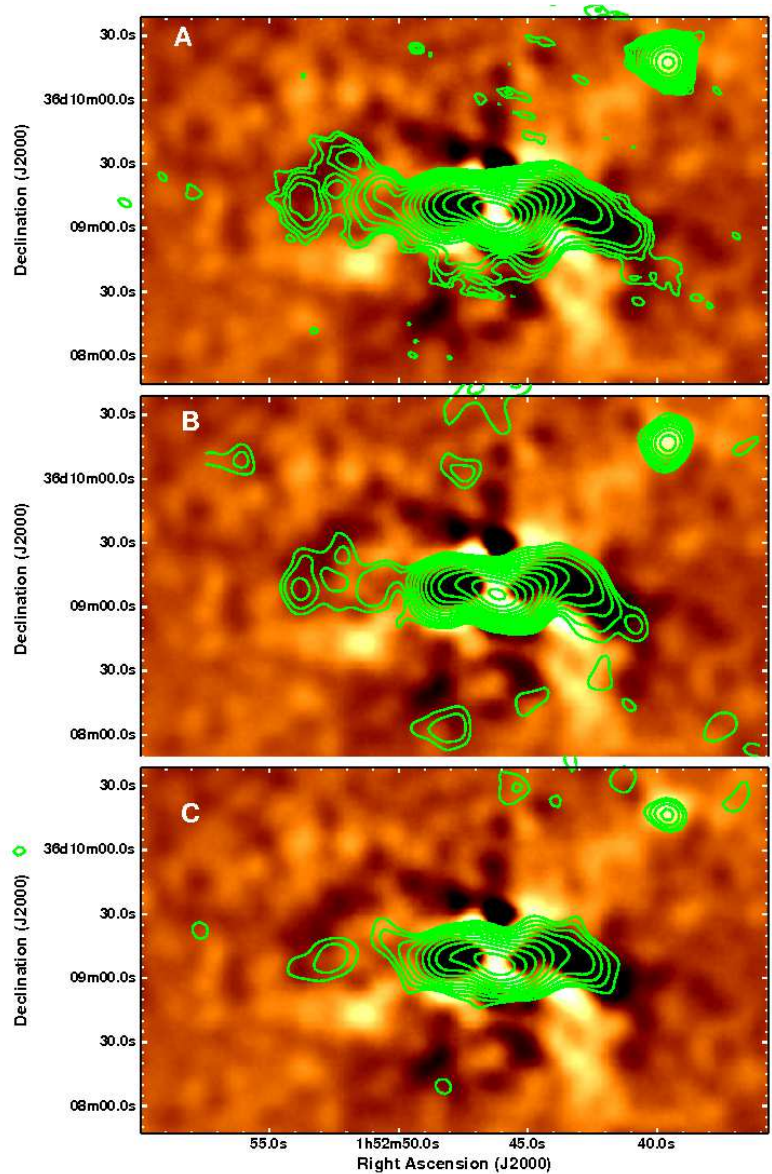}
\caption{{\footnotesize Residual X-ray image of Abell 262
    (Figure~\ref{fig:A262_unsharp}) with radio contours overlaid. {\bf
      A:} Tapered GMRT 610 MHz data where the data have been weighted
    to produce an image at lower resolution ($\theta$=13\farcs1
    $\times$ 13\farcs1) that is more sensitive to lower surface
    brightness emission. The radio contours increase by $\sqrt{2}$
    from 3$\sigma$, where $\sigma$=0.12 mJy/bm. The radio emission
    traces the full length of the southwest tunnel and shows evidence
    of a series of connected clumps to the east which are coincident
    with the newly detected extended eastern deficit. {\bf B:} The radio
    contours show the VLA 330 MHz radio emission plotted for the two
    sigma contour and higher ($\sigma$=0.66 mJy/bm, $\theta$=13\farcs1
    $\times$ 13\farcs1). Although this data is not as sensitive as the
    610 MHz GMRT data it shows the same morphology of the radio
    emission along the tunnel as well as the clumps associated with
    the extended eastern deficit. {\bf C:} The GMRT 235 MHz radio
    emission is less sensitive than the other frequencies but still
    shows the radio emission tracing the length of the southwest
    tunnel and some clumpy emission in the southern part of the
    eastern deficit. The resolution at 235 MHz is $\theta$=13\farcs1
    $\times$ 13\farcs1 and the contours increase by $\sqrt{2}$ from
    3$\sigma$, where $\sigma$= 1.1 mJy/bm.
\label{fig:mos2}} }
\end{figure}

We have also analyzed the 610 MHz data applying a suitable taper to
the $uv$ distribution to produce a map sensitive to the large scale
diffuse emission. The taper was chosen to allow us to match the
13\farcs1 beam of our lowest frequency observation at 235 MHz. The
tapered 610 MHz image (Figure~\ref{fig:mos2}A) reveals radio emission
extending the full length of the tunnel and shows a series of
connected clumps of emission running $\sim$ 22 kpc eastward (in
projection) from the eastern X-ray cavity. These previously unknown
radio clumps are coincident with the newly detected extended outer
eastern cavity discussed in Section~\ref{sect:chan_anal}. An inspection of
the DSS images shows that there are no apparent optical counterparts
to the eastern emission regions. The total (projected) linear extent
of the radio emission at 610 MHz is $\sim$ 60 kpc, more than 3 times
larger than previously reported by \citet{Parma86}. Similar evidence
of larger source sizes at low frequencies has been seen in other
cluster-center radio sources such as M87 \citep{owen00}, Perseus
\citep{fabian02} and Hydra A \citep{lane04} where the large outer
structures have been interpreted as signatures of one or more past AGN
outbursts. The clumpy morphology of the eastern emission and total
extent of the radio emission in Abell 262 is also seen in
Figure~\ref{fig:mos2}B where we show the VLA 330 MHz data convolved to
the same resolution as the tapered 610 MHz data.

Our lowest frequency observations at 235 MHz are shown in
Figure~\ref{fig:mos2}C. These data also trace the western radio
emission over the full length of the X-ray tunnel. On the eastern side
of the source, the 235 MHz observations reveal radio emission located
well beyond the eastern compact source out to the edge of the 610 MHz
extensions seen in Figure~\ref{fig:mos2}A. At the 3$\sigma$ contour
level this emission does not show a continuous connection to B2
0149+35 but this is likely due to the lower sensitivity of the 235 MHz
data.

\subsection{Spectral Index}
\label{sect:spectral}

\begin{figure}[tb]
\epsscale{1.0}
\plotone{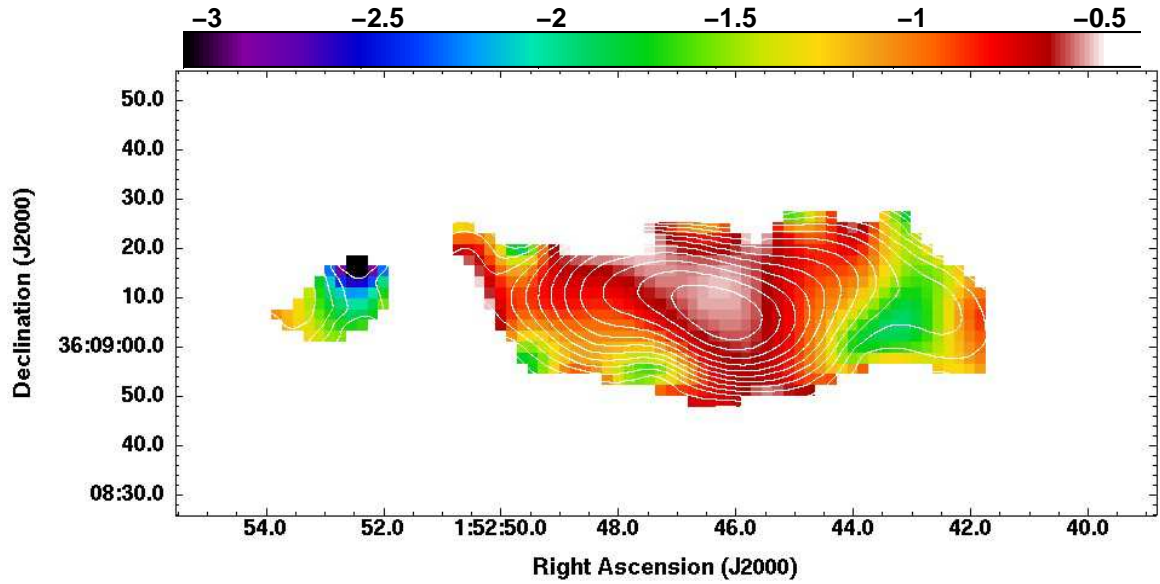}
\caption{Spectral index map between 235 and 610 MHz. The 610 MHz total
  intensity contours are overplotted to help connect the spectral
  properties to the morphology. The core shows a relatively flat
  spectral index ($\alpha=-0.5$). The spectrum shows significant
  steepening away from the core to the west as well as a smaller
  spectral steepening away from the core to the east.
\label{fig:spix}} 
\end{figure}

We have made a spectral index map of the central radio source using
the 235 and 610 MHz GMRT data. The $uv$-coverage of both data sets was
matched and both frequencies were imaged with a circular 13\farcs1
$\times$ 13\farcs1 beam. We show this spectral index map in
Figure~\ref{fig:spix} where we have blanked the spectral index map at
the 3$\sigma$ level on each of the input maps. The colorbar at the top
shows the (linear) spectral index scale running from $\alpha = -3.0$
to $\alpha = -0.4$, where $S_\nu \propto \nu^\alpha$. We have overlaid
the 610 MHz tapered contours in white on Figure~\ref{fig:spix} to
provide a reference for the source morphology in the locations where
we have measured the spectral index. Typical uncertainties in the flux
measurements due to calibration and systematic errors at the GMRT are
of the order of 5\% at 235 and 610 MHz \citep{Lal06}, resulting in a
typical error on the spectral index of $\sigma_\alpha$=0.2.

The spectral index in the area of the radio core is fairly flat
($\alpha = -0.5$) which is typical of recently injected particles. The
spectral index toward the western part of the source steepens away
from the core reaching a value of $\alpha = -1.7$ in the region of the
X-ray tunnel. The integrated spectral index of the radio emission in
the tunnel is $\alpha_{235}^{610}=-1.3\pm 0.2$. The origin of the
apparent spectral flattening of the emission on the western edge of
the tunnel is uncertain and may be a result of low signal to noise at
the edge of the 235 MHz image since the apparent steepening is within
roughly the 3$\sigma$ errors on the spectral index map. Toward the
eastern edge of the source the spectral index steepens less
dramatically away from the core reaching $\alpha_{235}^{610}=-0.9$. The region of the eastern
radio clumps shows an even steeper spectral index of roughly
$\alpha_{235}^{610}=-1.8$. Although we cannot use the current data to
trace small spectral index variations accurately across the source, we
note that the spectral steeping observed from the core toward the
outer source edges is significant above the 3$\sigma$ level and likely
traces spectral aging of the relativistic particles as they travel
further from the core. We have also made spectral index maps using the
330 MHz and 1400 MHz images and find similar evidence of spectral
steepening away from the core in those maps.

\subsection{Integrated Spectrum}
\label{sect:int_spec}

\begin{figure}[tb]
\plotone{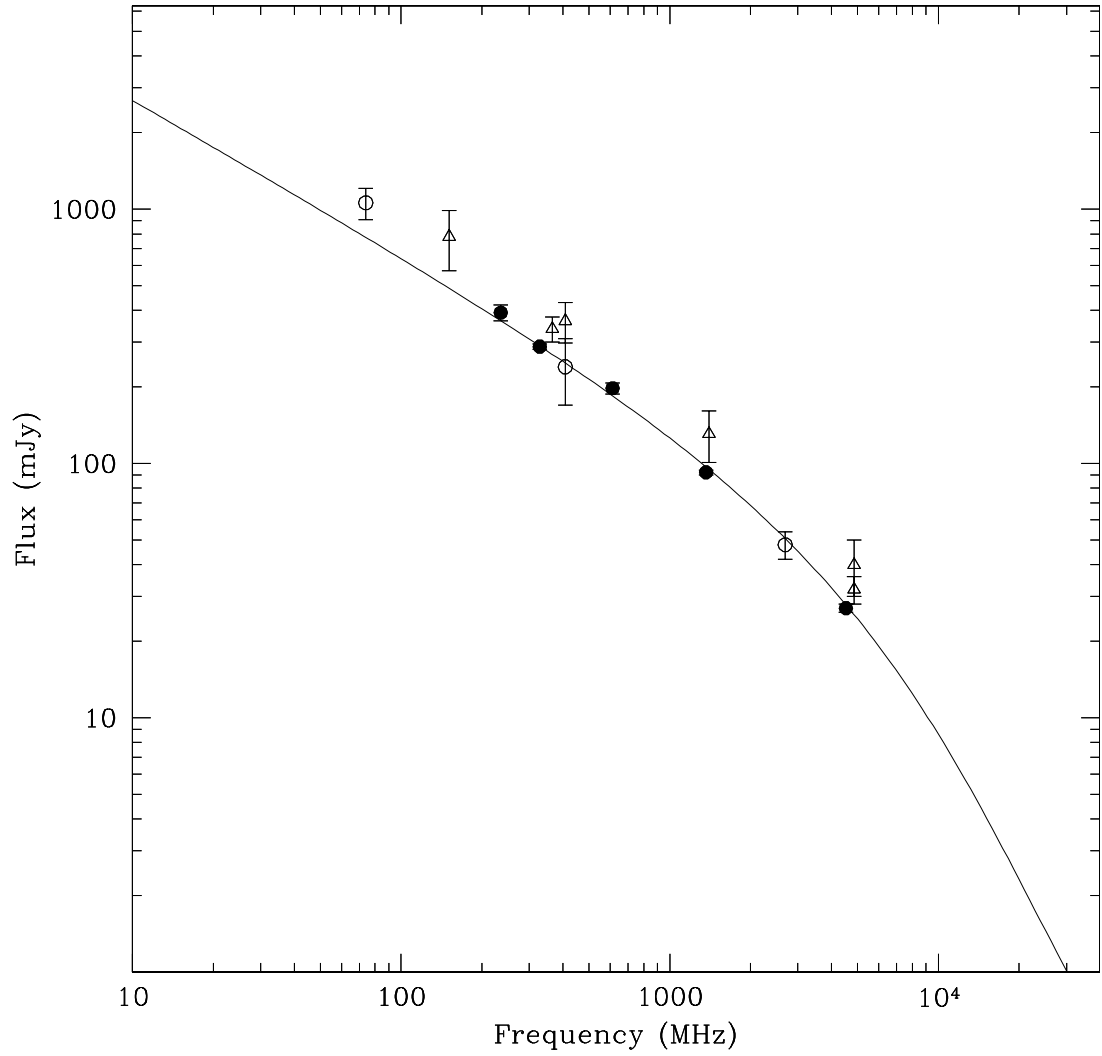}
\caption{Integrated radio flux measurements for Abell 262 from this
paper as well as the literature. Filled circles show the data points
from this work and open symbols represent measurements taken from the
literature. The open circles are from observations with sufficient
resolution to separate the emission of the central radio source from
the neighboring sources while the open triangles are low resolution
observations which likely contain contamination from the nearby
sources. The solid line shows the best-fit KP spectral model.
\label{fig:spectrum}} 
\end{figure}

The spectral index determined for the full source measured from our
observations between 235 and 4535 MHz is $\alpha=-0.90\pm 0.16$, where
we have assumed an uncertainty of 5\% on the GMRT 235 MHz flux, and
1\% on the VLA 4535 MHz flux. Assuming a similar flux uncertainty of
5\% on the GMRT 610 MHz observations we find an integrated low
frequency spectral index of $\alpha=-0.71\pm 0.17$ between 235 and 610
MHz. These observations suggest a possible flattening of the radio
spectrum to low frequencies but we note that this cannot be confirmed
from these data due to the relatively large errors on the GMRT flux
estimates. We have searched the literature for total flux measurements
of B2 1049+35 in order to determine the overall spectral shape of the
radio emission. We list the available flux measurements in
Table~\ref{tbl:fluxes} and plot the radio spectrum including both our
new measurements and the data from the literature in
Figure~\ref{fig:spectrum}. Several of the measurements from the
literature were obtained with sufficiently large beams that the
measured flux is contaminated by sources surrounding B2 0149+35. These
measurements are marked with triangles in the figure.

The shape of the radio spectrum can be used to estimate the age of the
source by using models to fit the balance of energy injection and
losses. Although the available flux measurements for B2 0149+35 cover
roughly a factor of 65 in frequency, they still do not sample a large
portion of the radio spectrum. We have fit all measurements in
Table~\ref{tbl:fluxes} to several particle injection models to
determine the break frequency\footnote{The break frequency is the
  frequency above which the synchrotron emission spectrum steepens due
  to particle losses.} of the emission of the full source. The fits
were done for both the full sample of flux measurements and also the
subset of uncontaminated flux measurements. The three models used to
fit the data were the JP (Jaffe-Perola), KP (Kardashev-Pacholczyk) and
CI (continuous injection) spectral models. The JP and KP models both
assume a single ``one-shot'' injection of a power-law distribution of
electrons with no further acceleration \citep{JP,K,P}. The JP model
allows for pitch-angle scattering, resulting in an exponential cutoff
in the emission spectrum, while the KP model does not allow pitch-angle
scattering and thus has an emission spectrum consisting of two
power-laws. The CI model assumes a continual injection of a power-law
distribution of relativistic particles resulting in a spectral break
that changes by $-0.5$ in $\alpha$. All three models produce similar
fits to the data and predict the break frequency lies in the range of
1400 to 9600 MHz. We find similar spectral break frequencies if we
include or exclude the contaminated flux measurements from the
literature. We have also separately fit the break frequency for the
bright emission filling the eastern cavity as well as the fainter
emission in the tunnel and to the east of the X-ray cavity. These
regions were fit using only the data presented in this paper as the
individual regions needed to be isolated to determine the fluxes. The
break frequency in the bright regions is between 2200 and 12000 MHz
while in the fainter extended emission it is between 1900 and 7400
MHz. For all models we assume an initial injection spectrum of
$\alpha=-0.6$ which is the measured spectral index of the source
between 330 and 610 MHz.

The average radiative age of the synchrotron emission is given by
\citet{Miley80} as:
\begin{equation}
\tau=0.82 \frac{B^{0.5}}{B^2 + B_{IC}^2} [(1+z)\nu_{b}]^{-0.5}\ \  {\rm yr}
\end{equation}
where $B$ is the magnetic field strength in the source in Gauss,
$B_{IC}$ is the equivalent magnetic field strength of the microwave
background [$4\times 10^{-6} (1+z)^2$ Gauss], and $\nu_{b}$ is the
break frequency in GHz. Assuming a break frequency at the low end of
the model predictions of 1400 MHz and that the source field strength is
$B=B_{IC}/\sqrt{3}$ to give the maximum radiative lifetime \citep{vp69}, we
estimate the age of B2 0149+35 as $\sim 47$ Myr. If the break
frequency is near the high end from the spectral fits (9600 MHz), we
then obtain a younger age of $\sim$ 18 Myr.

\section{Radio Lobe Interaction with the ICM}

The central radio source in Abell 262 appears to have had a
significant impact on the thermal gas distribution in the center of
this cluster. \citet{Blanton04} showed the cluster core contains
several knots of X-ray emission as well as at least one X-ray cavity
to the east of the radio core while our new residual $Chandra$ image
reveals the presence of an X-ray tunnel running westward from the AGN
as well as an extended outer eastern cavity
(Figure~\ref{fig:A262_unsharp}). We use our observations of the
synchrotron emission to obtain estimates of the radio properties for
comparison with the surrounding thermal gas properties.

\subsection{ICM Pressure Balance}
\label{sect:pressure}

The western extension along the X-ray tunnel is fairly faint at 1400
MHz and thus the spectral index of the extension is not well
constrained at high frequencies. In \S~\ref{sect:spectral} we estimate
the low frequency spectral index of the emission in the tunnel to be
$\alpha_{235}^{610}=-1.3\pm 0.2$. We use this spectral index
measurement together with the flux measurement at 610 MHz to estimate
the minimum energy properties of the western radio extension. We
follow the notation of \citet{Miley80} and assume that the emission
from the western extension comes from a uniform prolate cylinder with
a filling factor ($f$) of unity and that there is equal energy in
relativistic ions and electrons ($\kappa=1$). We use a source size of
$\sim 16.5 \times 6.0$ kpc and a model where the magnetic field is
perpendicular to the line of sight and the spectral index is constant
between the lower cutoff frequency of 10 MHz and the upper frequency
cutoff of 100 GHz. Using these parameters for the tunnel, we calculate
a minimum energy magnetic field strength of $B_{me}$ = 5.0 $\mu$G, a
non-thermal pressure of $P_{me} = 1.4\times 10^{-12}$ dyn cm$^{-2}$
and a synchrotron lifetime at 1400 MHz of $\tau_{me} = 37$ Myr. Note
that adiabatic expansion reduces the loss rate of the relativistic
electrons, and as a result the synchrotron lifetime may be shorter
than derived from the observed radio properties.

The X-ray gas pressure in the region surrounding the tunnel is
estimated from the $Chandra$ observations to be $P_{X}=6.4\times
10^{-11}$ dyn cm$^{-2}$. This X-ray pressure is more than 25 times
larger than our estimated minimum energy synchrotron pressure. This
discrepancy between the synchrotron and X-ray pressure is typical of
many cooling core systems as shown by \citet{dunn05}. Similar pressure
discrepancies are found in the case of lobes associated with several
more powerful FRII sources as well \citep[e.g.][]{hw}. Such a pressure
difference cannot be physical since it would result in a rapid
collapse of the radio lobes. This may suggest that the assumptions
used in our minimum energy calculations are wrong. Equipartition
between the synchrotron plasma and thermal gas would require
$\kappa/f$=260, which would suggest that the lobes are fed by heavy
jets. An alternative solution to the pressure discrepancy is that
there is an additional form of pressure support in the lobes (e.g.\ a
very hot diffuse thermal gas).

The synchrotron emission coincident with the outer eastern X-ray
cavity has a spectral index ranging from $\alpha=-1.0$ to as steep as
$\alpha= -1.5$ between 235 and 610 MHz. Using the flatter spectral
index and the minimum energy assumptions above, we estimate a minimum
energy magnetic field strength of $B_{me}$ = 2.6 $\mu$G, a non-thermal
pressure of $P_{me} = 4.1\times 10^{-13}$ dyn cm$^{-2}$ and a
synchrotron lifetime (neglecting adiabatic expansion) at 1400 MHz of
$\tau_{me} = 46$ Myr. We estimate an average X-ray pressure of
$P_{X}=5\times 10^{-11}$ dyn cm$^{-2}$ in the same region as the
synchrotron emission. As discussed above, this pressure difference
cannot be physical and likely indicates either incorrect minimum
energy assumptions or the presence of an additional form of pressure
support within these regions.

\subsection{Buoyant Rise Time in the X-ray Tunnel}

If the western radio extension represents a buoyant lobe from a
previous outburst then we can estimate the buoyancy rise time for the
lobe to reach the end of the tunnel.  The sound speed in the thermal
gas is $c_s=\sqrt{\frac{\gamma kT}{\mu m_{H}}}$ where $\gamma=5/3$ is
the adiabatic index of the gas, $T$ is the X-ray temperature of the
gas, $\mu$ is the mean molecular weight of the gas, and $m_H$ is the
mass of the proton. Using $T$= 1keV from \citet{Blanton04}, the sound
speed in the central cluster region is $c_s= 475$ km s$^{-1}$. As the
lobe rises it will approach a terminal velocity that is determined by
the balance of buoyancy and drag forces. Based on the terminal
velocity approach of \citet{churazov}, \citet{mn07} give the terminal
velocity in terms of the Kepler speed ($v_K$) as $v_b \propto
(4v_k/3)(2r/R)^{0.5}$, where $r$ is the bubble radius, $R$ is the bubble
distance from the cluster center and the Kepler speed is given by
$v_K$=$(gR)^{0.5}$. The local gravitational acceleration, $g$, can be
estimated from the local stellar velocity dispersion assuming the
galaxy is an isothermal sphere, $g\simeq 2\sigma^2/R$. The stellar
velocity dispersion for NGC 708 is given by \citet{bern02} as 235 km
s$^{-1}$, which results in a terminal velocity of $v_b=$ 245 km
s$^{-1}$ for the outer bubble.

Based on the above terminal velocity, we find a buoyant lobe would
reach the observed (projected) location at the end of the tunnel on a
timescale $\tau_{buoy} \gsim 80$ Myr. Note that the lower limit on the
buoyancy timescale is due to projection since the radio structure may
not be in the plane of the sky. On the other hand, if the tunnel
remains evacuated of thermal ICM between successive outbursts then the
rise time calculated above overestimates the timescale for the
synchrotron emitting plasma to reach the end of the tunnel. We cannot
make a firm statement regarding the need for particle re-acceleration
in the tunnel due to the uncertainties in the estimates of the
synchrotron lifetime of the particles in the tunnel
(\S~\ref{sect:pressure}) as well as the uncertainties in the buoyancy
rise time.

\subsection{AGN Outburst Timescale}
\label{sect:timescale}

The radio emission on the eastern side of the core (beyond the eastern
lobe) likely represents emission from one or more previous outburst
episodes from the central AGN. We measure projected distances from the
center of the eastern cavity of $\sim$ 10, 17, and 21 kpc for the
three eastern structures. Assuming that each of these represents a
buoyant lobe from a past AGN outburst and using the calculated
terminal velocity of the buoyant bubbles ($v_b=$ 245 km s$^{-1}$), we
can estimate the outburst repetition rate in this system to be
$\tau_{rep}\sim 28$ Myr\footnote{Note that we chose to use the distance from the center
of the eastern cavity for three outbursts rather than the cluster core
and four outbursts since it is unlikely that the inner active lobe has
been rising at the terminal velocity while this is likely a good
approximation for the outer detached lobes.}. This outburst timescale is similar to the
estimate for Perseus using the observed spacing of the X-ray ripples
\citep{fabian03}.  Such a short repetition
timescale would suggest that multiple outbursts may be responsible for
creating the emission seen in the tunnel. We note that it seems
unlikely that the repetition rate is as long as the typically assumed
$10^8$ yr since buoyancy arguments would require the bubbles to be
aligned at an inclination of less than 20\degr\ from the line of
sight. An alternative interpretation for the new radio morphology is
that the three eastern clumps may represent a much larger fragmented
bubble from a single outburst that occurred prior to the current
activity powering the inner lobes. In this case (even without
considering projection effects) our repetition timescale would be an
underestimate of the true repetition timescale.

\subsection{Energy Balance: Total AGN Input vs Cooling Luminosity}
\label{sect:ebal}
Our new radio observations of B2 0149+35 show significant structure in
the radio source to the east of the X-ray cavity. Interpreting the
synchrotron emission as a series of buoyantly rising bubbles from past
radio outbursts allows us to obtain a lower limit on the total energy
input into the ICM over at least four outburst episodes in this
source. We assume that the bubbles rise buoyantly and undergo
adiabatic expansion in pressure equilibrium with the surrounding
thermal gas. Following the notation of \citet{churazov02}, the initial
energy input from the outburst is related to the final bubble
properties by
\begin{equation}
E_o=\frac{\gamma}{\gamma-1}p_fV_f \left( \frac{p_o}{p_f}\right) ^\frac{\gamma -1}{\gamma}
\end{equation}
where $\gamma$ is the adiabatic index of the plasma in the lobes,
$p_o$ is the initial ambient pressure where the lobe was originally
inflated, $p_f$ is the surrounding pressure at the current location of
the buoyant lobe, and $V_f$ is the volume of the bubble at the current
location. Due to sensitivity limitations, X-ray depressions are
difficult to isolate for all radio features of B2 0149+35 and thus the
cavity volumes are estimated based on the observed synchrotron
emission seen at 610 MHz. The radio emission is less sensitive to
projection effects than X-ray depressions \citep[as also noted
  by][]{birzan08}.

\begin{figure}[tb]
\plotone{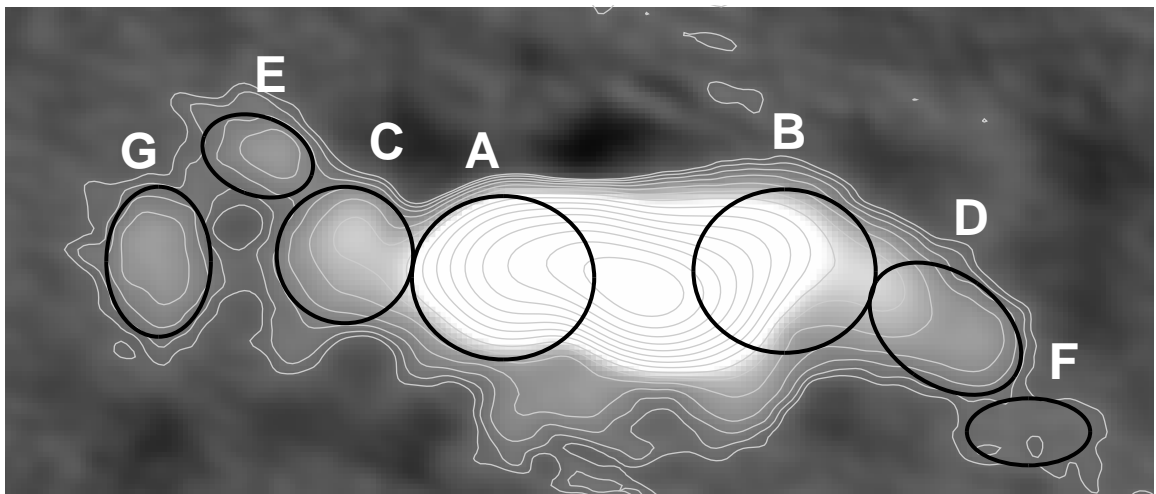}
\caption{Tapered GMRT 610 MHz image with individual source components
  shown as overlaid black ellipses. Each component was identified
  based on the radio morphology of the source.
\label{fig:comps}} 
\end{figure}

The A262 cavity system provides us with the opportunity to estimate
variations in outburst energies over several episodes for the same
AGN. We have separated the radio source into 7 components (4 east of
the core and 3 west of the core; Figure~\ref{fig:comps}) and
calculated the initial input energy from the AGN required to create
these regions. The inner (spherical) eastern cavity was defined based
on the X-ray cavity size and morphology. We use the same bubble shape
and volume to match the inner western cavity while the remainder of
the western X-ray tunnel is represented by a prolate cylinder. At the
end of the western tunnel we define a small prolate cylinder based on
the radio morphology. The outer eastern components were defined based
on the radio morphology visible above the 5$\sigma$ level. The first
component is spherical while the outer two components are represented
by prolate cylinders. The bubbles were assumed to be initially
inflated at a radius consistent with that of the active eastern radio
lobe and associated X-ray cavity. Using the location of the active
eastern cavity rather than the cluster center as the inflation and
detachment point avoids complications from likely supersonic or
transonic jets closer to the cluster core. The X-ray pressure
surrounding each region was determined from the radial profile
obtained from the $Chandra$ observations.

In Table~\ref{tbl:energies} we present a summary of the sizes and
initial input energies calculated for each component. The lower limit on
the initial energy input varies from $\sim 5\times 10^{56}$ ergs up to
$\sim 8\times 10^{57}$ ergs, while the total energy summed over all 7
radio components is $2.2\times 10^{58}$ ergs. This total energy input
is similar to the average energy calculated for typical `single'
outburst systems \citep{birzan}. From the results in
Table~\ref{tbl:energies} we find that the current epoch of activity
from B2 0149+35 is the most energetic if our underlying assumptions on
the origin of the radio morphology hold. In fact, the distribution of
the outburst energy for each of the components of the source shows
that the energy appears to be roughly consistent with a continual
increase to the current epoch. Such an increase in outburst energy may
be the signature of a system where the cooling luminosity continues to
dominate over the AGN energy injected into the ICM leading to an
increase in the fueling of the AGN with time and thus more energetic
outbursts. The observed trend in outburst energy for B2 0149+35 is
different from that seen for the multi-cavity system in Hydra A
\citep{wise07}. In that case, \citet{wise07} find that the energy
required to create the outer cavities is nearly an order of magnitude
larger than that for the inner cavities. We also note that the cavity
size trend for B2 0149+35 seen in Table~\ref{tbl:energies} does not
follow the bubble size versus radius trend seen in the cavity sample
of \citet{diehl08}. 

Based on the radio morphology, we suggest that the source components
are the result of 4 outburst episodes that have a repetition time of
$\sim28$ Myr. The total AGN kinetic luminosity over these outbursts is
$L_{AGN}=6.2\times 10^{42}$ ergs s$^{-1}$. \citet{Blanton04} estimate
the cooling luminosity in A262 to be $L_{cool}=1.3\times 10^{43}$ ergs
s$^{-1}$, which is roughly a factor of two higher than the estimated
AGN kinetic luminosity. Although \citeauthor{Blanton04} showed that
the current radio outburst is apparently too weak to offset cooling
assuming $\tau_{rep}=10^8$ yr, our new data suggest that the AGN in
NGC 708 can offset cooling on average with our shorter repetition
timescale over the observed multiple outburst episodes. If the eastern
radio morphology is instead the result of fragmentation from a single
outburst episode it is likely that the energy output estimate above
would underestimate the true AGN output.

\subsection{Radiative Efficiency}

The presence of the intracluster medium confining the radio lobes
allows us to compare estimates of the kinetic energy of the radio jets
to the observed bolometric radio luminosity of the system. The
radiative efficiency determined from embedded AGNs provides important
insights into the impact of AGN outbursts on the evolution of the host
galaxy and the intracluster medium as well as information relating to
AGN accretion and jet formation models. Theoretical estimates of the
radiative efficiency fall in the range of $5-10$\%
\citep{deyoung93,bicknell95}. An observational estimate of the
radiative efficiency of 10\% was obtained by \citet{wold07} using a
sample of FRI and FRII sources from the 3CRR catalog of
\citet{liang83}. Using a sample of radio sources associated with
cluster-center cavities, \citet{birzan} find the radiative
efficiencies fall in the range of a few tenths of a percent to a few
percent.

The total radio luminosity of B2 0149+35 was calculated by integrating
the flux between 10 MHz and 10 GHz assuming a constant spectral index
of $\alpha=-0.90$ and the total flux as measured from our 235 MHz
observations. We find the bolometric radiative power to be $L_{radio}=
3.9\times 10^{39}$ ergs s$^{-1}$, roughly 50\% higher than the
bolometric luminosity calculated in \citet{birzan08}. A lower limit on
the total AGN energy output is obtained from the mechanical power
deposited into the ICM from this radio source ($pV$ work) divided by
the timescale for the energy input. From \S~\ref{sect:ebal} we
estimate the total AGN kinetic luminosity to be $L_{AGN}=6.2\times
10^{42}$ ergs s$^{-1}$, giving a radiative efficiency similar to
\citet{birzan} of $\sim$ 0.1\%, reinforcing the view that this source
appears to be a very inefficient radiator.

\subsection{Mass Accretion Rate}

The inferred black hole mass for B2 0149+35 can be calculated using
the velocity dispersion versus black hole mass relation of
\citet{tremaine02}. The stellar velocity dispersion of 235 km s$^{-1}$
for NGC 708 \citep{bern02} corresponds to a black hole mass of
$2.6\times 10^8 M_\odot$. The total energy output estimated for the
four outburst episodes traced by the radio emission implies an average
accretion rate of 0.001 $M_\odot$ yr$^{-1}$ assuming $\epsilon =0.1$
as the efficiency of converting accreted mass to kinetic energy
outflow. This mass accretion estimate is similar to that determined by
\citet{r06} from the analysis of only the inner X-ray cavities in
Abell 262.

\section{Summary}

We have undertaken a detailed multi-frequency radio study of B2
0149+35 which is associated with the cluster-center galaxy in Abell
262. Our new low frequency radio observations reveal that the source
is nearly three times larger than previously known. The S-shaped
source seen by \citet{Parma86} is surrounded by emission to the east
and west. The western emission extends from the edge of the
\citeauthor{Parma86} source toward the southwest. A comparison of high
and low resolution images shows that this emission appears to be
composed of several synchrotron clumps which are surrounded by diffuse
emission. The eastern extension beyond the \citeauthor{Parma86} source
is composed of three clumps of emission which are connected to the
main source by a faint synchrotron bridge. Spectral index images of
the emission reveal a steepening of the spectrum away from the radio
core.

A morphological comparison of the radio and X-ray emission in Abell
262 shows a significant interaction in the cluster core. The innermost
of the newly identified eastern radio structures falls at the position
of a low significance X-ray deficit identified by
\citet{Blanton04}. The other two outer eastern radio features appear
to be associated with a newly discovered X-ray depression. The western
radio emission completely fills the newly discovered X-ray tunnel seen
in our residual $Chandra$ image. Similar X-ray tunnels filled with low
frequency emission are seen in Abell 2597 \citep{clarke05} and Perseus
\citep{fabian06}. These tunnels may represent regions where multiple
radio outbursts pile up and accumulate over several AGN activity
cycles.

The integrated radio spectrum of B2 0149+35 shows evidence of spectral
steepening to high frequencies. Using three particle injection models
(Jaffe-Perola, Karsashev-Pacholczyk, and continuous injection), we
estimate the break frequency to be between 1400 and 9600 MHz. Under
the source field strength assumption given in
Section~\ref{sect:int_spec}, this range of break frequencies leads to
upper limits on the radiative age of the source of between 47 and 18
Myr. Minimum energy synchrotron arguments yield a synchrotron lifetime
of $\sim$ 37 Myr for the source at frequencies around 1400 MHz. These
timescales can be compared to the buoyancy timescale ($\tau_{buoy}
\gsim$ 80 Myr) calculated assuming that the western lobe is from a
past outburst and that it detached and rose buoyantly to the end of
the X-ray tunnel. From this comparison, it is not clear whether active
particle injection or particle acceleration is required within the
tunnel to maintain the observed radio emission due to the large
uncertainties in the synchrotron age (\S 4.1) and buoyancy timescale
(\S 4.2). We have also used the projected separation of the eastern
clumps of emission in the buoyancy model (assuming the lobes rise at
the terminal velocity) to estimate a lower limit of $\tau_{rep}$=28
Myr on the average repetition rate of outbursts for the central AGN.

The X-ray pressure surrounding the extended radio emission is
estimated to be more than 25 times larger than the minimum energy
synchrotron pressure calculated if we assume a synchrotron filling
factor of unity and equal energy in relativistic ions and
electrons. This discrepancy may suggest that the radio lobes are fed
by heavy, baryon-dominated plasma. Alternatively there may be an
additional form of pressure support within the radio lobes.

The total energy input into the ICM of Abell 262 over multiple
outburst episodes was calculated by assuming that each radio component
to the east of the core represents a lobe from an AGN outburst. In
this model the western emission filling the tunnel would be the result
of a series of overlapping buoyant lobes. The total energy input over
the 4 outbursts is estimated to be $2.2\times 10^{58}$ ergs. By
comparison, the outburst in clusters such as Abell 2052
\citep{Blanton09} and Perseus \citep{fabian00} ranges from a few times
10$^{58}$ to a few times 10$^{59}$ ergs. The most powerful known radio
outburst is in MS0735.6+7421 with an estimated energy of $6\times
10^{61}$ ergs \citep{mcnamara05}.

We compared the lower limit on the AGN energy input into the ICM to
the bolometric radio luminosity of B2 0149+35 to estimate a radiative
efficiency of $\epsilon=0.001$. This efficiency agrees well with that
previously calculated for this system by \citet{birzan}.

Although previous measurements of the kinetic radio power of B2
0149+35 indicated that the current outburst was too weak to offset the
radiative cooling, our new measurements suggest that the source can
approximately offset the radiative cooling on average over the four
observed outburst episodes.

The ubiquitous nature of low frequency radio emission as a chronicle
of recent and past AGN activity in clusters underscores the importance
of emerging instruments such as LOFAR and the LWA for delineating the
role of AGN feedback in the ICM.
 
\acknowledgements

The National Radio Astronomy Observatory (NRAO) is a facility of the
National Science Foundation operated under cooperative agreement by
Associated Universities, Inc.  We thank the staff of the GMRT that
made these observations possible, and Samir Dhurde in particular for
his assistance. GMRT is run by the National Centre for Radio
Astrophysics of the Tata Institute of Fundamental Research. The
Digitized Sky Surveys were produced at the Space Telescope Science
Institute under U.S. Government grant NAG W-2166. The images of these
surveys are based on photographic data obtained using the Oschin
Schmidt Telescope on Palomar Mountain and the UK Schmidt
Telescope. Basic research in radio astronomy at the NRL is supported
by 6.1 Base funding. TEC was supported in part by the National
Aeronautics and Space Administration through $Chandra$ awards
GO6-7115B, and GO7-8132B.  ELB was supported in part by the National
Aeronautics and Space Administration through $Chandra$ award G07-8132A
and a Clare Boothe Luce Professorship. CLS was supported in part by
the National Aeronautics and Space Administration through $Chandra$
awards GO5-6081X, GO5-6126X, and AR7-8012X.

\clearpage

\begin{deluxetable}{lccccc}
\tablecaption{Radio Observations of B2 0149+35}
\tablewidth{0pt}
\tablehead{
\colhead{Date} & \colhead{Instrument/Array}   & \colhead{Frequency}   &
\colhead{Bandwidth} &
\colhead{Duration} &
\colhead{Obs.\ Code}\\
\colhead{} & \colhead{} & \colhead{(MHz)} & \colhead{(MHz)} & \colhead{(hours)} & \colhead{}
}
\startdata
1997 Sep 18 & VLA/C & 1364.9/1435.1 & 50/50 & 0.6 & AC488\\
2000 Oct 01 & VLA/D & 4535.1 & 50 & 0.4 & AC557\\
2000 Dec 19 & VLA/A & 1365/1415.5 & 25/25 & 0.4 & AC572\\
2000 Dec 19 & VLA/A & 1465/1515 & 25/25 & 0.4 & AC572\\
2003 Jul 27  & VLA/A & 321.5/328.5 & 6.3/6.3 & 3.5 & AC682\\
2004 Jul 24  & GMRT & 240 & 8.0 & 8.5 & 06TEC01\\
2004 Jul 24  & GMRT & 606/622 & 16/16 & 8.5 & 06TEC01\\
2005 Apr 16 & VLA/B & 1396.5 & 6.3 & 1.7 & AM828\\
\enddata
\label{tbl:radio_obs}
\end{deluxetable}

\clearpage

\begin{deluxetable}{rcc}
\tablecaption{Flux Measurements}
\tablewidth{0pt}
\tablehead{
\colhead{Frequency} & \colhead{Flux} & \colhead{Reference}\\
\colhead{(MHz)} & \colhead{(mJy)} & \colhead{}}
\startdata
74 & 1060$\pm$140 & 1\\
151 & 780$\pm$ 207 & 2\\
235 & 392$\pm$ 20 & 3\\
365 & 339$\pm$ 38 & 4\\
328 & 288$\pm$ 4 & 3\\
408 & 364$\pm$ 66 & 5\\
408 & 240$\pm$ 70 & 6\\
613 & 198$\pm$ 10 & 3\\
1365 & 92$\pm$ 2 & 3\\
1400 & 131$\pm$ 30 & 7\\
2695 & 48$\pm$ 6 & 8\\
4535 & 27$\pm$ 1 & 3\\
4850 & 40$\pm$ 10 & 9\\
4850 & 32$\pm$ 4 & 10\\
\enddata
\label{tbl:fluxes}
\tablerefs{
(1) Cohen et al.\ 2006; (2) Hales et al.\ 1993; (3) This work;
(4) Douglas et al.\ 1996; (5) Colla et al.\ 1973; (6) Ballarati et al.\ (1983); (7) White \& Becker 1992;
(8) Rudnick \& Owen 1977; (9) Andernach et al.\ 1980; (10) Gregory et al.\ 1996.}
\end{deluxetable}

\clearpage
\begin{deluxetable}{crrrr}
\tablecaption{Bubble Energies}
\tablewidth{0pt}
\tablehead{
\colhead{Bubble} & \colhead{a} & \colhead{b} & \colhead{r} & \colhead{E$_o$}\\
\colhead{} & \colhead{(kpc)} & \colhead{(kpc)} & \colhead{(kpc)} & \colhead{($10^{56}$ ergs)}\\
\colhead{(1)} & \colhead{(2)} & \colhead{(3)} & \colhead{(4)} & \colhead{(5)}}
\startdata
A &  5.56 & 4.99 & 8 & 82\\
B &  5.56 & 4.99 & 10 & 68\\
C &  4.16 & 4.16 & 18 & 25\\
D &  5.10 & 3.37 & 20 & 20\\
E &  3.48 & 2.39 & 25 & 6.1\\
F &  3.76 & 2.04 & 26 & 4.7\\
G &  4.58 & 3.19 & 29 & 13\\
\enddata
\label{tbl:energies}
\tablecomments{ Col.(1): Bubble ID as indicated in
Figure~\ref{fig:comps}. Col.(2): Projected semimajor axis of
bubble. Col.(3) Projected semiminor axis of bubble. Col.(4)
Bubble-center projected radial distance from the core of
B2~0149+35. Col.(5) Initial AGN outburst energy required to create bubble.}
\end{deluxetable}

\clearpage

\clearpage

\clearpage

\clearpage

\clearpage

\clearpage

\clearpage

\end{document}